\documentclass[pra,reprint,showpacs,amsmath,amssymb,superscriptaddress,longbibli
ography,eqsecnum]{revtex4-1}

\usepackage{bm,color,graphicx,braket,mathptmx,amssymb}
\usepackage[colorlinks=true,linkcolor=blue,citecolor=blue,urlcolor=blue]{hyperref}

\newcommand{\Tr}{\mathrm{Tr}}

\begin{document}

\title{Optimal measurements for quantum spatial superresolution} 

\author{J. \v{R}eh\'a\v{c}ek} 
\affiliation{Department of Optics,
 Palack\'y University, 17. listopadu 12, 771 46 Olomouc, 
Czech Republic}

\author{Z. Hradil}
\affiliation{Department of Optics,
 Palack\'y University, 17. listopadu 12, 771 46 Olomouc, 
Czech Republic}

\author{D. Koutn\'{y}}
\affiliation {Department of Optics,
 Palack\'y University, 17. listopadu 12, 771 46 Olomouc, 
Czech Republic} 

\author{J. Grover} 
\affiliation{ESA---Advanced ConceptsTeam, European Space Research
  Technology Centre (ESTEC), Keplerlaan 1, Postbus 299, NL-2200AG
  Noordwijk, Netherlands}

\author{A. Krzic} 
\affiliation{ESA---Advanced ConceptsTeam, European Space Research
  Technology Centre (ESTEC), Keplerlaan 1, Postbus 299, NL-2200AG
  Noordwijk, Netherlands}

\author{L. L. S\'{a}nchez-Soto} 
\affiliation{Departamento de \'Optica,
  Facultad de F\'{\i}sica, Universidad Complutense, 
28040~Madrid,  Spain} 
\affiliation{Max-Planck-Institut f\"ur die Physik des Lichts,
  Staudtstra{\ss}e 2, 91058 Erlangen,
  Germany}

\begin{abstract}
  We construct optimal measurements, achieving the ultimate precision
  predicted by quantum theory, for the simultaneous estimation of
  centroid, separation, and relative intensities of two incoherent
  point sources using a linear optical system. We discuss the physical
  feasibility of the scheme, which could pave the way for future
  practical implementations of quantum inspired imaging. 
\end{abstract}

\maketitle

\section{Introduction}

Metrology is the science of devising schemes that extract as precise
as possible an estimate of the parameters associated with a system.
The quantum foundations of this field were laid years
ago~\cite{Helstrom:1976ij,Holevo:2003fv}; since then, most of the
efforts have been devoted to single-parameter estimation, with a
special emphasis in the prominent example of
phase~\cite{Giovannetti:2011aa,Demkowicz:2015aa}. The quantum
Cram\'er-Rao lower bound (qCRLB) then provides a saturable bound on
the estimation uncertainty and recipes for finding the optimal
measurement attaining that limit are known~\cite{Braunstein:1994aa}.

The case of multiparameter estimation is considerably more
involved~\cite{Yuen:1973aa,Belavkin:1976aa,Szczykulska:2016aa}. Although
the equivalent qCRLB was formulated long time
ago~\cite{Helstrom:1967aa}, this bound is not always saturable.  The
intuitive reason for this is the incompatibility of the measurements
for different parameters. The conditions under which the qCRLB can be
saturated have been determined~\cite{Matsumoto:2002aa,Gill:2013aa} The
associated optimal measurements have been worked out for pure
states~\cite{Pezze:2017aa}, but for mixed states the results are
fragmentary~\cite{Yue:2014aa,Zuppardo:2015aa,Proctor:2018aa}.

In this work we will address these problems in the context of the
two-point resolution limit for an optical system.  In classical optics
several criteria exist~\cite{Dekker:1997aa,Hemmer:2012aa,
  Cremer:2013aa} to quantitatively determine these limits, the most
famous of which is due to Rayleigh~\cite{Rayleigh:1879aa}.

Most of these criteria exploit properties of the point spread function
(PSF) that specifies the intensity response to a point light source.
This provides an intuitive picture of the mechanisms limiting
resolution, but also has several shortcomings. These mainly stem from
the fact that these criteria were developed for the human eye as the
main detector.  For example, the Rayleigh limit is defined as the
distance from the center to the first minimum of the PSF, which can be
made arbitrarily small with ordinary linear optics, although at the
expense of the side lobes becoming much higher than the central
maximum~\cite{Schermelleh:2010aa}. This confirms that determining the
position of the two points becomes also a question of photon
statistics rather than being solely described by the Rayleigh limit.

A careful reconsideration of this conundrum has been performed in the
framework of quantum estimation theory~\cite{Tsang:2016aa,Nair:2016aa,
  Ang:2016aa,Lupo:2016aa,Rehacek:2017aa,Kerviche:2017aa,
  Tsang:2018aa,Zhou:2018aa}. This work showed that, in the case of two
identical incoherent point sources with \emph{a priori} knowledge of
their centroid, the precision of an optimal measurement stays constant
at all separations. As a consequence, the Rayleigh limit is subsidiary
to the problem and arises because standard direct imaging
discards all the phase information contained in the field. These
predictions fuelled a number of proof-of-principle
experiments~\cite{Paur:2016aa,Yang:2016aa, Tham:2016aa,Yang:2017aa}.

While remarkable, this result does not hold in the more general case
of two unequally bright sources. In a suitable multiparameter
scenario~\cite{Chrostowski:2017aa}, where simultaneous estimation of
centroid, separation, and relative brightness was considered, it was
found that their estimation precisions decreased with
separation~\cite{Rehacek:2017ab}.  Nonetheless, an appropriate
strategy was shown to lead to a significant improvement in precision
at small separations over direct imaging for any fixed number of
photons.  The measurements attaining the ultimate
quantum limits for this case are relevant to a number of applications, for
example observational astronomy and microscopy.

\section{Model and associated multiparameter quantum 
Cram\'er-Rao bound} 

To be as self-contained as possible, we first set the stage for our
analysis. We assume a linear spatially invariant system illuminated
with quasimonochromatic paraxial waves with one specified
polarization. We consider one spatial dimension; $x$ denoting the
image-plane coordinate.

We phrase what follows in a quantum language that will simplify the
following calculations.  To a field of complex amplitude $U( x )$ we
assign a ket $| U \rangle $, such that $U( x )= \langle x | 
U \rangle$, $| {x} \rangle$ being a point-like
source at ${x}$. The system PSF is denoted by
$I (x) = | \langle x | \Psi \rangle |^{2} = |\Psi (x)|^{2}$, so that
$\Psi(x)$ can be interpreted as the amplitude PSF.

Two point sources, of different intensities and separated by a
distance $\mathfrak{s}$, are imaged by that system. Since they are
incoherent with respect to each other, the total signal must be
depicted as a density operator 
\begin{equation}
  \varrho_{\bm{\theta}}= \mathfrak{q} \, \varrho_{+} 
  + (1-\mathfrak{q} ) \, \varrho_{-} \, ,
\end{equation}
where $\mathfrak{q}$ and $1 - \mathfrak{q}$ are the intensities of the
sources (the total intensity is normalized to unity).  The individual
components $\varrho_{\pm}= |\Psi_{\pm}\rangle \langle\Psi_{\pm}|$ are
just $x$-displaced PSF states; that is,
$ \langle x | \Psi_{\pm} \rangle = \langle x - \mathfrak{s}_{0} \mp
\mathfrak{s}/2 | \Psi \rangle$, so that they are symmetrically located
around the geometric centroid $\mathfrak{s}_{0}$. Note that
\begin{equation}
  | \Psi_{\pm} \rangle = \exp[ - i (\mathfrak{s}_{0} \pm
  \mathfrak{s}/2) P ]  | \Psi \rangle \ ,
\end{equation}
where $P$ is the momentum operator, which generates displacements in
the $x$ variable, and acts as a derivative
$P \mapsto - i \partial_{x}$.

The measured density matrix depends on the centroid $\mathfrak{s}_{0}$,
the separation $\mathfrak{s}$, and the relative intensities of the
sources $\mathfrak{q}$. This is indicated by the vector
$\bm{\theta} = (\mathfrak{s}_{0}, \mathfrak{s}, \mathfrak{q})^{t}$.
Our task is to estimate the values of $\bm{\theta}$ through the
measurement of some observables on $\varrho_{\bm{\theta}}$.

In this multiparameter estimation scenario, the central quantity is
the quantum Fisher information matrix (qFIM)~\cite{Petz:2011aa}. This
is a natural generalization of the classical Fisher information, which
is a mathematical measure of the sensitivity of a quantity
to changes in its underlying parameters. However, the qFIM is
optimized over all the possible measurements. It is defined as
\begin{equation}
  Q_{\alpha \beta} (\bm{\theta}) = \frac{1}{2} 
  \Tr ( \varrho_{\bm{\theta}} \{ L_{\alpha}, L_{\beta} \}) \, ,
\end{equation} 
where the Greek indices run over the components of the vector
$\bm{\theta}$ and $\{ \cdot, \cdot \}$ denotes the anticommutator.
Here, $L_{\alpha}$ stands for the symmetric logarithmic
derivative (SLD)~\cite{Helstrom:1967aa} with respect the parameter
$\theta_{\alpha}$:
  \begin{equation}
    \frac{1}{2} (L_{\alpha} \varrho_{\bm{\theta}} +
    \varrho_{\bm{\theta}} L_{\alpha} ) = \partial_{\alpha}
    \varrho_{\bm{\theta}} \, ,
  \end{equation}
with $\partial_{\alpha} = \partial/ \partial \theta_{\alpha}$.

The qFIM is a distinguishability metric on the space of quantum states
and leads to the multiparameter qCRLB~\cite{Braunstein:1994aa,Szczykulska:2016aa} 
for a single detection event:
\begin{equation}
  \label{singleevent}
  \mathrm{Cov} (\widehat{\bm{\theta}} ) \ge
  Q^{-1}  (\bm{\theta} )  \, ,
\end{equation}
where $\mathrm{Cov} (\widehat{\bm{\theta}})$ is the covariance matrix
for a locally unbiased estimator $\widehat{\bm{\theta}}$ of the
quantity $\bm{\theta}$. Its matrix elements are
$\mathrm{Cov}_{\alpha\beta} (\widehat{\bm{\theta}} ) = \mathbb{E} [
(\widehat{\theta}_{\alpha} - \theta_{\alpha})
(\widehat{\theta}_{\beta} - \theta_{\beta}) ]$, $\mathbb{E} [Y]$ being
the expectation value of the random variable $Y$.  The above
inequality should be understood as a matrix inequality. In general, we
can write $\Tr [ \mathcal{C} \, \mathrm{Cov} (\widehat{\bm{\theta}} )
]  \ge \Tr [ \mathcal{C} \, Q^{-1} (\bm{\theta} ) ] $, where $\mathcal{C}$ is
some positive cost matrix, which allows us to asymmetrically
prioritise the uncertainty cost of different parameters.

Unlike for a single parameter, the collective bound in
(\ref{singleevent}) is not always saturable, as the measurements for
different parameters may be incompatible~\cite{Holevo:2003fv}. 
 The multiparameter qCRLB can be saturated provided
\begin{equation}
\Tr ( \varrho_{\bm{\theta}}   [ L_{\alpha},  L_{\beta} ]  ) = 0 \, ,
 \label{eq:cond}
\end{equation}
where $[ \cdot , \cdot ]$ is the commutator. This condition is
necessary and sufficient for pure
states~\cite{Matsumoto:2002aa,Gill:2013aa}, upon which the criterion
is equivalent to the existence of some pair of SLDs that commute. It
is then possible to find an optimal measurement as the common
eigenbasis of these SLDs. For mixed states, this criterion has been
discussed by a number of authors~\cite{Suzuki:2016aa} and has met some
small inconsistencies in its usage, being variously identified as
sufficient~\cite{Vaneph:2013aa} or necessary and
sufficient~\cite{Crowley:2014aa}.  Reference~\cite{Ragy:2016aa} offers
a clear account of this question.  For our particular case,
\eqref{eq:cond} is fulfilled whenever the PSF amplitude is
real~\cite{Rehacek:2017ab}, $\Psi(x)^\ast=\Psi(x)$, which will be
assumed henceforth ensuring that the parameters are therefore
compatible.

For the model we are considering, and after a lengthy
calculation~\cite{Rehacek:2017ab}, we obtain a compact expression for
the qFIM; viz
\begin{equation}
  \label{qfi}
  Q = 4 
  \left(
    \begin{array}{ccc}
      p^{2} + 4 \mathfrak{q} (1- \mathfrak{q}) \wp^{2} 
      & (\mathfrak{q}-1/2) p^{2}
      & -i w  \wp \\
      ( \mathfrak{q}-1/2) p^{2} & p^{2}/4 & 0\\
      -i w \wp & 0 & \displaystyle  \frac{1-w^2}
                     {4 \mathfrak{q}(1- \mathfrak{q})}
    \end{array}
  \right) \, , 
\end{equation}
which depends solely on the quantities
\begin{eqnarray}
  w & \equiv & \langle \Psi_{\pm} |\Psi_{\mp} \rangle = 
               \langle \Psi |\exp(i \mathfrak{s} P) |\Psi \rangle \, , 
               \nonumber  \\ 
  p^{2} & \equiv & \langle \Psi_{\pm} |P^2 |\Psi_{\pm}\rangle 
                   = \langle \Psi | P^2 | \Psi\rangle \, ,  \\
  \wp & \equiv& \pm \langle \Psi_{\pm} | P| \Psi_{\mp}\rangle =
                \langle \Psi|\exp(i \mathfrak{s} P) P|\Psi\rangle \, . \nonumber 
\end{eqnarray}
The quantity $p^{2}$ is determined by the shape of the PSF, whereas
both $w$ and $\wp$ (which is purely imaginary) depend on the
separation $\mathfrak{s}$.

Only for equally bright sources, $\mathfrak{q}=1/2$, the measurement
of $\mathfrak{s}$ is uncorrelated with the other parameters.  In
general, when $\mathfrak{q} \neq 1/2$ the separation is correlated
with the centroid (via the intensity term $\mathfrak{q}-1/2$) and the
centroid is correlated with the intensity (via $p^{2}$).

The individual parameter $\theta_{\alpha}$ can be estimated with a
variance satisfying
$\mathrm{Var} (\widehat{\theta}_{\alpha}) \ge (Q^{-1})_{\alpha \alpha}
(\bm{\theta} )$.  It is convenient to use the inverses of the
variances $ {H}_{\alpha} = 1/\mathrm{Var} (\theta_\alpha)$, usually
called the precisions~\cite{Bernardo:2000aa}.  By inverting the QFIM
and taking the limit $\mathfrak{s} \rightarrow 0$, they turn out to
be~\cite{Rehacek:2017ab}
\begin{eqnarray}
  \label{approx}
    H_{\mathfrak{s}_{0}}^{Q} &  \simeq & \mathcal{Q}^{2}\,
    G_{22} \,  \mathfrak{s}^{2} + O(\mathfrak{s}^4) \, ,\nonumber \\
    H _{\mathfrak{s}}^{Q} &  \simeq & \frac{\mathcal{Q}^{2}}
    {4(1-\mathcal{Q}^{2})} \, G_{22}
    \, \mathfrak{s}^2 + O(\mathfrak{s}^4) \, ,\\
    H_{\mathfrak{q}}^{Q} &  \simeq  & \frac{1}{\mathcal{Q}^{2}} G_{22} \,
    \mathfrak{s}^4 + O(\mathfrak{s}^6)\, , \nonumber 
\end{eqnarray}
where
\begin{eqnarray}
\label{eq:refsu}
\mathcal{Q}^2& = & 4\mathfrak{q}(1-\mathfrak{q})< 1  \,, \nonumber \\
& & \\
G_{22}^{2} & = & \mathrm{Var}(P^2)=\langle\Psi |P^4 |\Psi\rangle-
\langle\Psi| P^2 |\Psi\rangle^{2} \, . \nonumber
\end{eqnarray}
The superscript $Q$ indicates that
the quantities are evaluated from the quantum matrix $Q$.

\section{Optimal measurements}

We shall focus on finding measurements attaining the quantum limit,
thus offering significant advantages with respect to conventional
direct intensity measurements. In the general case of unequally bright
sources ($\mathfrak{q}\neq 1/2$), the lack of symmetry makes this
issue challenging and one cannot expect to find closed-form
expressions for the optimal positive operator valued measures (POVMs)
for all the values of the source parameters. However, this becomes
viable when separations get very small. As already discussed, this is
the most interesting  regime, where conventional imaging techniques
fail. 

We start by  specifying a basis in the signal space. A suitable choice
is the set $\{| \Psi_{n} \rangle \} $ defined in terms of the spatial
derivatives of the amplitude PSF:
\begin{equation}
  \langle x|\Psi_n\rangle=\frac{\partial^n}{\partial x^n} 
  \Psi(x-x_{0}),
  \quad n=0,1,2,\ldots,
\end{equation}
where $x_{0}$ is an arbitrary displacement in the
$x$-representation. We convert this set into an orthonormal
basis $\{ | \Phi_{n} \rangle \}$ by the standard Gram-Schmidt process.
In this basis, all results can be expressed in a PSF-independent
form.  Moreover, signals well centered on the origin and with small
separation, are represented by low-dimensional states; i.e.,
$\varrho_{\bm{\theta}} \rightarrow |\Phi_0 \rangle \langle \Phi_0|$
for $\mathfrak{s}_0\rightarrow x_{0}$, and
$\mathfrak{s}\rightarrow 0$.

To estimate three independent parameters, the required POVM must have
at least four elements. We therefore consider the following class of
measurements $\Pi_{j} = | \pi_j \rangle \langle \pi_j|$,
$j= 0, \ldots, 2$ and
$\Pi_{3} = \openone - \Pi_{0} - \Pi_{1} - \Pi_{2}$, so only three of
these are independent.  The first three POVM elements are defined in a
four-dimensional subspace, with basis
$\{ |\Phi_0\rangle, \ldots, |\Phi_{3}\rangle\}$, wherein we expand
$| \pi_{j}\rangle$ ($j=0, \ldots, 2$) as
\begin{equation}
  \label{meas}
  |  \pi_j\rangle= \sum_{k=0}^3 C_{j k}|\Phi_k\rangle \, .
\end{equation}
Obviously, the projectors $| \pi_j \rangle \langle \pi_j|$ must be
linearly independent.  In addition, we impose the following set of
conditions
\begin{equation}
\label{conditions}
\begin{array}{lcc}
& |\Phi_0\rangle & |\Phi_1\rangle\\
& & \\
|\pi_0\rangle \quad & C_{00}=0 \quad & C_{01}\neq 0 \\
|\pi_1\rangle \quad & C_{10}=0 \quad & C_{11} \neq 0\\
|\pi_2\rangle \quad & C_{20}\neq 0 \quad & C_{21} \neq 0 \, ,
\end{array}
\end{equation}
where the row index can be permuted. In this way, two of the three
rank-one projectors are orthogonal to the signal PSF $\Phi_0$; a
crucial factor boosting the performance of the measurement. We stress
that, by changing the displacement $x_{0}$, the basis and the
measurement itself is displaced.

Next, we expand the signal components in the small parameter. 
We define $a_\pm= \mathfrak{s}_0 \pm  \mathfrak{s}-x_{0}$,  so we have 
$\langle x|\Psi_\pm\rangle= \Psi(x-x_0-a_\pm)$, and the expansion in 
$a_{\pm}$ gives 
\begin{eqnarray}
\langle
x|\Psi_\pm\rangle& = & \sum_m\frac{(-a_{\pm})^{m}}{m!} 
 \frac{\partial^m}{\partial x^m}\Psi(x-x_0) = 
 \sum_m\frac{(-a_{\pm} )^m}{m!}\langle x|\Psi_m\rangle \nonumber \\
&=  &
\langle x|\sum_n |\Phi_n\rangle \sum_m \frac{(-a_{\pm})^m}{m!} G_{nm}
\,, 
\end{eqnarray}
where $G_{nm} = \langle \Phi_n|\Psi_m\rangle$ [note that
  $G_{22}$ in Eq.~\eqref{eq:refsu} is consistent with this general
  definition].  Keeping terms up to the fourth power, we get
\begin{eqnarray}
    |\Psi_\pm\rangle & \simeq &  
\left(G_{0 0}+   \frac{a_\pm^2}{2}   G_{02} + 
  \frac{a_\pm^4}{24} G_{04}\right)|\Phi_0\rangle   \nonumber \\
&  + &  \left(a_\pm G_{11}+\frac{a_\pm^3}{6} G_{13}\right)| \Phi_1\rangle
       \nonumber \\
    & + & \left(\frac{a_\pm^2}{2} G_{22}+\frac{a_\pm^4}{24}
      G_{24}\right)|\Phi_2\rangle  \nonumber \\
& + & \frac{a_\pm^3}{6}G_{33}|\Phi_3\rangle \, .
\end{eqnarray}
Notice that for real amplitude PSFs, all $G$s carrying both odd and
even subscripts are zero. This follows from the fact that
$\langle \Psi_n |\Psi_m\rangle=0$, and hence
$\langle \Psi|P^{m+n}|\Psi\rangle=0$ for any combination of odd and
even subscripts, whenever the wave-function is real. We also have
$G_{nm}=0$ for all $n>m$, by construction of the basis set, which
makes a basis function orthogonal to all lower-order non-orthogonal
functions, as the latter span a subspace that the former is
orthogonal to.

We are set to evaluate the probabilities
\begin{equation}
  p_j=\mathfrak{q}\langle \Psi_+|\Pi_j|\Psi_+\rangle +
  (1-\mathfrak{q}) \langle\Psi_-|\Pi_j|\Psi_- \rangle \, ,
\end{equation} 
and the corresponding classical Fisher information matrix per
detection event:
\begin{equation}
  \label{fisher}
  F_{\alpha\beta}=\sum_{j=0}^{3} 
  \frac{(\partial_\alpha  p_j) (\partial_\beta p_j)}{p_j} \, .
\end{equation}
The maximum of the classical Fisher information $F$ is its quantum
version $Q$, as $Q$ is optimized over all POVMs. The corresponding
precisions are thus related by $H_\alpha^{Q} \ge H_\alpha$.

Our initial strategy is to align the center of the
measurement~(\ref{meas}) with the signal centroid by letting 
$x_{0}=\mathfrak{s}_0$. The calculation  of  the precisions  turns out
to be a very lengthy task, yet the final result is surprisingly
simple
\begin{equation}
  \label{measprec}
  H_{\alpha}= \lambda H_{\alpha}^{Q} \, .
\end{equation}
Therefore,  $H_{\alpha}$ differs from the quantum limit precision 
by a factor
\begin{equation}
  \label{factor}
  \lambda = \mathcal{Q}^2 \mathcal{A}\, , \qquad 
  \mathcal{A}=\frac{(C_{01} C_{12}-C_{02}
    C_{11})^2}{C_{01}^2+C_{11}^2}< 1 \, .
\end{equation}
The coefficient $\lambda$ consists of the product of two factors: one
depending solely on the intensities [as defined in
Eq.~\eqref{eq:refsu}], the other depending on the measurement. The
latter one will be called the quality factor of the measurement.
Conditions \eqref{conditions} are crucial for deriving relations
\eqref{measprec} and \eqref{factor}: violating them makes the dominant
terms of $H_\alpha$ disappear and kills the superresolution. One
pertinent example would be projection on the basis set
$|\Phi_k\rangle$: $C_{jk}=\delta_{jk}$ as for example projections on a
set of Hermite-Gauss modes for a Gauss PSF advocated in
Refs.~\cite{Tsang:2016aa} and \cite{Rehacek:2017aa} among others. Such
projections can be optimal for estimating separation, but ultimately
fail when separation, centroid, and intensity are to be estimated
together in a multiparameter scenario considered here.

Going back to our result, two remarks are in order here. First, the
performance of the measurement (\ref{meas}), when aligned with the
centroid, scales with the same power of $\mathfrak{s}$ as the quantum
limit does. The quantum limit is attained, but for a separation
independent factor. This is true for all real-valued PSFs, no matter
how we set the remaining free parameters of the measurement. Second,
by optimizing those free parameters, the separation-independent factor
$\lambda$ can be made arbitrarily close to
$\lambda_{\mathrm{max}} = \mathcal{Q}^2$. Hence, for balanced signals
($\mathfrak{q}=1/2$), $\lambda_{\mathrm{max}} \rightarrow 1$ and the
measurement (\ref{meas}) becomes optimal. Conversely, for unbalanced
signals, the measurement is suboptimal and its performance worsens
with $\mathfrak{q}$, approaching the limit $\lambda \rightarrow 0$
when $\mathfrak{q}\rightarrow 0$ and $ \mathfrak{q}\rightarrow 1$.

\begin{figure}
  \centerline{\includegraphics[width=\columnwidth]{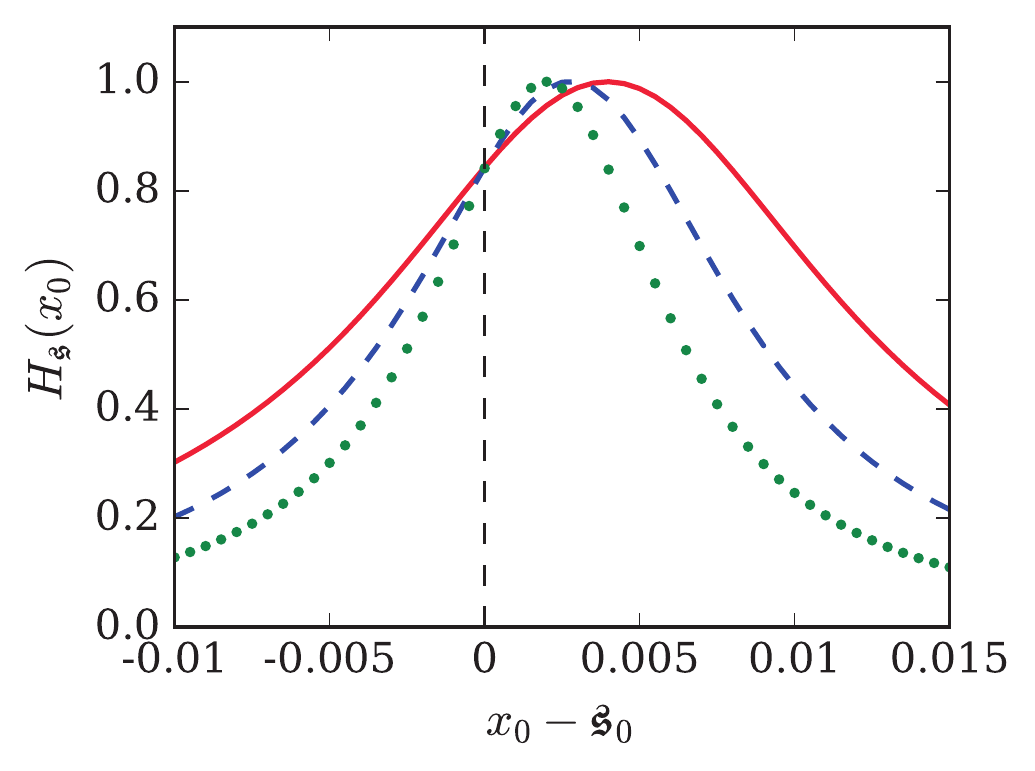}}
  \caption{The precision $H_{\mathfrak{s}}$ of the separation for
    point objects with relative intensities $\mathfrak{q}=0.3$ and
    $\mathfrak{s}=0.02$ (red solid line), $\mathfrak{s}=0.014$ (blue
    broken line), and $\mathfrak{s}=0.01$ (green dots) as a function
    of misalignment $x_{0} - \mathfrak{s_0}$ between the measurement
    displacement and the centroid. The maxima of the Lorentzians are
    normalized to unity to make the changes in widths and centers
    apparent.}
  \label{fig1}
\end{figure}

Next, we show that quantum limits can be saturated for any
$\mathfrak{q}$ by optimizing the displacement $x_{0}$. The
key point is that in the limit
$\mathfrak{s}\ll 1$, the precisions $H_\alpha (x_{0})$, when considered
as a function of the measurement displacement $x_{0}$, take a
Lorentzian shape, as can be appreciated in Fig.~\ref{fig1} for the
particular case of $H_{\mathfrak{s}} (x_{0})$.  On
decreasing the signal separation, the Lorentzian narrows down, with
its center approaching the signal centroid. We therefore adopt the
model 
\begin{equation}
  H_\mathfrak{s}(x_{0})= 
  \frac{\ell_{1} \mathfrak{s}^2}
  {1+ \displaystyle \frac{\ell_{2}(x_{0}-\mathfrak{s}_0+\ell_{3}\mathfrak{s})^2}
    {\mathfrak{s}^2}} \, .
\end{equation} 
The parameters can be identified by expanding $H_\mathfrak{s}$ in
$\mathfrak{s}$ and $x_{0} -\mathfrak{s}_0$:
\begin{equation}
\ell_{1} \mathfrak{s}^2 = \mathcal{A} H_\mathfrak{s}^{Q} \, , \quad
\ell_{2} = \frac{1}{\mathfrak{q}(1-\mathfrak{q})}  \,, \quad
\ell_{3}= \frac{1}{2} (1-2\mathfrak{q})\, .
\end{equation}
This uncovers the optimal  displacement and precisions
\begin{eqnarray} 
\label{displacement}
 &  x_{0}^{\mathrm{opt}}  =  
\arg  \max\limits_{x_{0}}H_\mathfrak{s}(x_{0})  =
\mathfrak{s}_{0}- \case{1}{2} \mathfrak{s}(1-2\mathfrak{q}) \, , &
\nonumber \\
& & \\
& H_\alpha(x_{0}^\mathrm{opt})  =   \mathcal{A} H_{\alpha}^{Q} \, . &
\nonumber
\end{eqnarray}
This is the central result of this paper. The optimal choice of
displacement is precisely
\begin{equation}
  x_{0}^{\mathrm{opt}} = (1- \mathfrak{q})
  (\mathfrak{s}_{0}-\mathfrak{s}/2)+\mathfrak{q}(\mathfrak{s}_{0}
  +\mathfrak{s}/ 2) \, ,
\end{equation} 
so that the weighted centroid, rather than the geometrical centroid,
is relevant to align the measurement.  Note that the weighted centroid
only coincides with the center of mass of the PSF when the PSFs are
symmetric.  By optimizing the measurement displacement $x_{0}$, the
intensity dependent $\mathcal{Q}^2$ term is removed from
(\ref{measprec}) and (\ref{factor}) and the qCRLBs are saturated for
all the signal parameters simply by letting
$\mathcal{A}\rightarrow 1$. As this can be done in infinitely many
ways, we conclude there are infinitely many measurements attaining the
quantum limit in multiparameter superresolution imaging. They can be
constructed following our recipe for any real-valued amplitude PSF.

\section{Examples}

To illustrate our result with a concrete example, we construct three
orthogonal vectors through 
\begin{equation}
  \label{optmeas}
  |\pi_{\,0,1}\rangle =
  \left ( 
    \begin{array}{c}
      0 \\ \frac{\sin(\phi/2)}{\sqrt{1+\cos\phi}} \\
  \pm\frac{\cos(\phi/2)}{\sqrt{1+\cos\phi}} \\
 -\sqrt{\frac{\cos\phi}{1+\cos\phi}}
    \end{array} 
  \right ) \, ,
  \quad
  |\pi_{2}\rangle=
  \left ( 
    \begin{array}{c}    
      \sqrt{\frac{2\cos\phi}{1+3\cos\phi}} \\ 
   \sqrt{\frac{2\cos\phi}{1+3\cos\phi}} \\
      0 \\ 
  \sqrt{\frac{1-\cos\phi}{1+3\cos\phi}}
    \end{array}
  \right ) 
\end{equation}
with $0<\phi<\pi/2$ in the $|\Phi_k\rangle$-representation, to build a
family of POVMs according to the recipe (\ref{meas}). This measurement
satisfies all the requirements, and the quality factor becomes
$ \mathcal{A}= 1$, so that the quantum limit is attained for any
real-valued PSF as long as $\mathfrak{s}\ll \sigma$.

\begin{figure}
 \centerline{\includegraphics[width=\columnwidth]{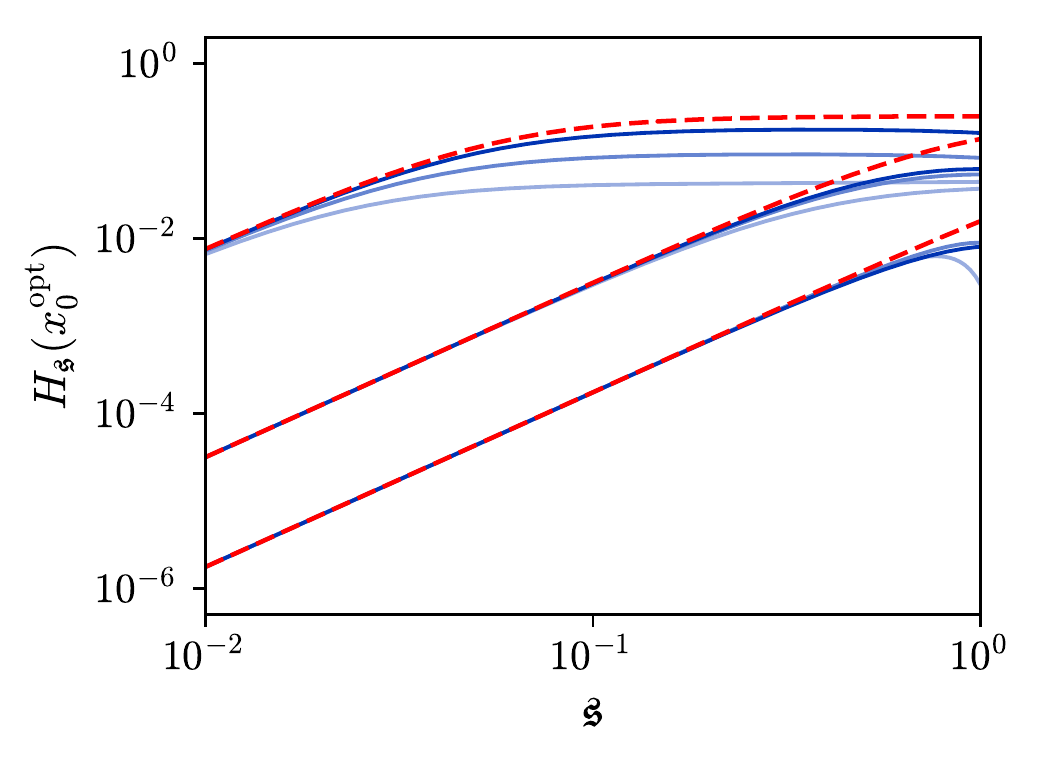}}
  \caption{The precision $H_{\mathfrak{s}} (x_{0})$ for an optimally displaced
    measurement (\ref{optmeas}) (blue lines) as compared to the
    quantum limit (\ref{approx}) (red broken lines). The lines are
    grouped by the intensity difference: $\mathfrak{q}=0.49$ (top),
    $\mathfrak{q}=0.35$ (middle), and $\mathfrak{q}=0.1$
    (bottom). Within each group (light to dark) $\phi=\pi/4$,
    $7\pi/20$, and $9\pi/20$, respectively. Notice the
    fast convergence towards the quantum limit as
    $\mathfrak{s}\rightarrow 0$.  A Gaussian PSF of a unit width
    $\sigma=1$ is assumed.} 
  \label{fig2} 
\end{figure}

The theory thus far is largely independent of the actual form of the
PSF. To be more specific we adopt a Gaussian PSF, with unit width
$\sigma=1$, which will serve from now on as our basis unit length.
The associated orthonormal basis is then a set of displaced
Hermite-Gauss modes
\begin{eqnarray}
  \Phi_{n} (x) &= & \frac{1}{(2\pi)^\frac{1}{4}2^\frac{n}{2}\sqrt{n!}}
  H_n\left ( (x-x_{0}^\mathrm{opt} ) /\sqrt{2}\right ) \nonumber \\
& \times &   \exp \left [ 
 - \tfrac{1}{4} (x-x_{0}^\mathrm{opt})^2
  \right ] \, ,
\end{eqnarray}
where $H_{n} (x)$ are the Hermite polynomials. In this case, we have
then $G_{22} = 1/8$. 

Figure~\ref{fig2} shows the resulting precision $H_\mathfrak{s}$ as a
function of~$\mathfrak{s}$ on a log-log scale for different
intensities $\mathfrak{q}$ and different measurements of the family
\eqref{optmeas}.  Direct numerical evaluation of the Fisher
information (\ref{fisher}) was done using a computational basis
$\{\Phi_{n}\}$ of dimension $30$ and no further approximation. With
$\mathfrak{s}\rightarrow 0$ all precisions quickly converge towards
the quantum limit and all the measurements (\ref{optmeas}) become
optimal. Notice however that performances over a wider range of
separations are sensitive to measurement parameter $\phi$ and values
close to $\phi=\pi/2$ provide the best overall performance.

Having potential applications of the new detection scheme in mind we
realize that achieving the quantum limits requires knowing the true
values of the measured parameters. In particular, the measurement must
be optimally displaced to reach the quantum limits and this
displacement, through (\ref{displacement}), depends on all the unknown
signal parameters. Consequently different displacements should be used
for different signals.

Can one hope to saturate the quantum limits for all signals with a
fixed measurement? Unfortunately, the answer is negative.  Let us
consider the estimation of a signal with strongly overlapping
components $\mathfrak{s}\ll 1$ of highly unequal intensities
$\mathfrak{q} \rightarrow 0$ (the same analysis can be carried out for
$\mathfrak{q} \rightarrow 1$), so that the weak component is
outshined. To gain significant information about the weak component,
the bright one must be almost completely suppressed in one of the
measurement outputs.  This is ensured by projecting the signal on a
state that is nearly orthogonal to the bright component. That crucial
projection, though, depends on both the signal centroid and
separation.

Our optimal measurement also behaves in this way. Let us look at the
value of $x_0^{\mathrm{opt}}$ in the limit
$\mathfrak{q}\rightarrow 0$; i.e., when $|\Psi_-\rangle$ is the bright
component.  In this case,
$x_0^\mathrm{opt} \rightarrow \mathfrak{s}_0-\mathfrak{s}$ coincides
with the center of the bright component. But, this means that
$|\Phi_0\rangle = |\Psi_-\rangle$ and the two outputs described by
$|\pi_0\rangle$ and $|\pi_1\rangle$ project on subspaces orthogonal to
the bright component, as anticipated.

\begin{figure}
  \centerline{\includegraphics[width=\columnwidth]{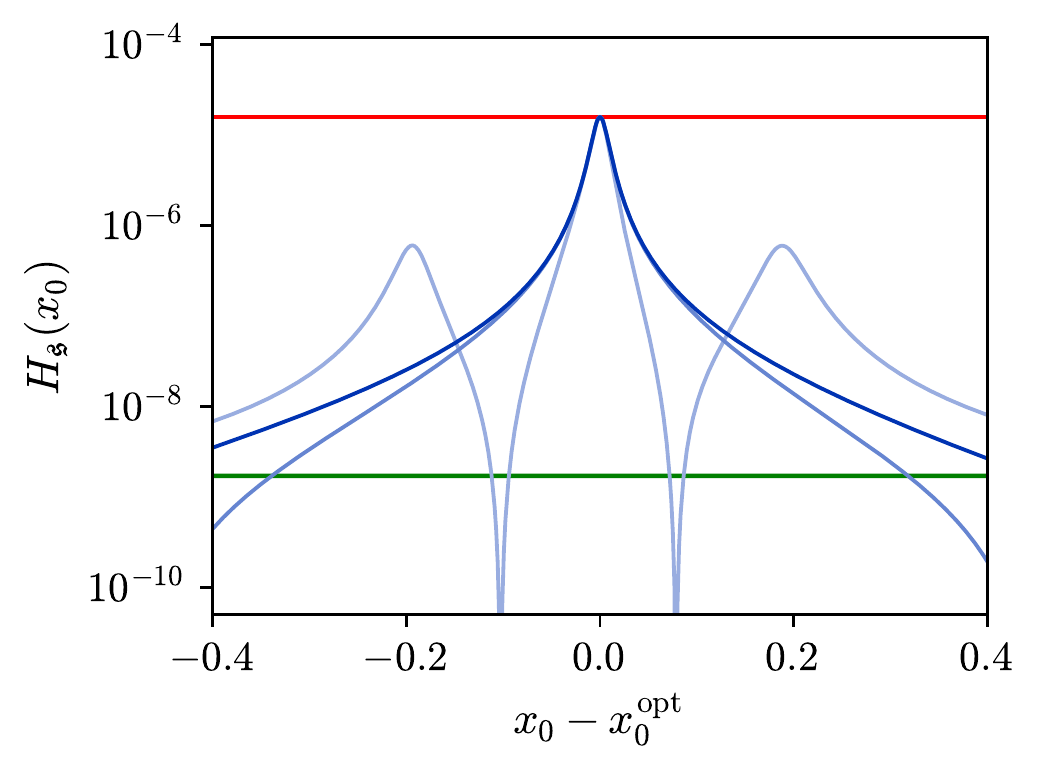}}
  \caption{The precision $H_{\mathfrak{s}}$ for misaligned
    measurements (\ref{optmeas}) (solid blue lines) compared to the
    corresponding quantum (top red) and direct imaging (bottom green)
    limits. The parameters of the measurements are $\phi=\pi/20$ (light
    blue), $\pi/2$ (light blue), and $9\pi/20$ (dark blue).  Observe
    the log vertical scale. A Gaussian PSF of a unit width $\sigma=1$ is
    assumed; and the sources satisfy $\mathfrak{s}=0.03$ and
    $\mathfrak{q}=0.1$.}
  \label{fig3}
\end{figure}

In practice, the performance will be compromised by any misalignment
with respect to $x_{0}^{\mathrm{opt}}$. This effect is examined in
Fig.~\ref{fig3}, where the quantum limit and the direct intensity
imaging are compared with different misaligned measurements
(\ref{optmeas}).  Being about two orders of magnitude below the
Rayleigh limit, such imperfections cause a loss of precision. Even
then, the advantage with respect to direct imaging persists over a
wide range of displacements $x_{0}$, demonstrating the robustness of
our detection scheme. Again, setting $\phi\approx\pi/2$ seems to be
the best option. For this particular example, the measurement can be
misaligned by as much as $0.4\sigma$ from $x_0^\mathrm{opt}$ and still
beat the direct imaging limits in measuring separations two orders of
magnitude below the Rayleigh limit.

Such an inherent robustness of optimal detection schemes hints at
using adaptive strategies to achieve the quantum limits. One plausible
way would be to spend a portion of the photon pool to obtain a first
estimate of the optimal displacement
$\hat{x}_0^\mathrm{opt}= \hat{\mathfrak{s}}_0 - \hat{\mathfrak{s}}(1
-2 \hat{\mathfrak{q}})/2$.  Since this quantity is closely related to
the weighted centroid, direct imaging can be used in this step. Then,
the estimated $\hat{x}_0^\mathrm{opt}$ can be used with the optimal
measurement \eqref{optmeas} in the next step to refine the estimates
of the signal parameters and so forth.

Having considered the fundamental aspects of the problem, how does
one implement the optimal measurement in practice for one particular
setting of the displacement?  This amounts to performing simultaneous
projections on three mutually orthogonal states. There
exists a unitary transformation taking this triplet into another set
of orthogonal vectors, where the latter set is experimentally
feasible. For example the optimal projections can be mapped on three
different pixels of a CCD camera. Unitary transformations of this
kind can be always realized with a set of non-absorbing
masks. Alternatively, giving up some performance, the implementation
can be facilitated by splitting the signal beam and measuring the three
projections separately. This leads to a photon loss and a threefold
decrease of the precisions $H_\alpha$, which can be tolerated for
sufficiently small separations.

\section{Conclusions}

We have examined the ultimate limits for the simultaneous estimation
of centroid, separation, and relative intensities of two incoherent
point sources. Our results indicate that the optimal sub-Rayleigh
resolution limit can be achieved for any real-valued amplitude PSF
provided the system output is projected onto a suitable complete set
of modes.  Particularly useful modes can be generated from the
derivatives of the system PSF, which in the limit of small separations
can access all available information with a few projections.

For equally bright sources, our proposed projection is optimal
whereas, for unbalanced signals, its performance deteriorates with the
parameter $\mathfrak{q}$. While some of our findings were illustrated
explicitly for Gaussian PSFs, our framework is general and can be
applied to other relevant cases.
 
All in all, this constitutes an important application of
multiparameter quantum estimation theory to a more realistic imaging
setting.  Our analysis provides a toolbox for achieving optimal
resolution and paves the way for further experimental demonstrations
and innovative solutions in scientific, industrial, and biomedical
domains.

\acknowledgments
We acknowledge financial support from the Grant Agency of the Czech
Republic (Grant No.~18-04291S), the Palack\'y University (Grant No.
IGA\_PrF\_2018\_003), the European Space Agency's
ARIADNA scheme, and the Spanish MINECO (Grant FIS2015-67963-P).


\begin{thebibliography}{40}%
\makeatletter
\providecommand \@ifxundefined [1]{%
 \@ifx{#1\undefined}
}%
\providecommand \@ifnum [1]{%
 \ifnum #1\expandafter \@firstoftwo
 \else \expandafter \@secondoftwo
 \fi
}%
\providecommand \@ifx [1]{%
 \ifx #1\expandafter \@firstoftwo
 \else \expandafter \@secondoftwo
 \fi
}%
\providecommand \natexlab [1]{#1}%
\providecommand \enquote  [1]{``#1''}%
\providecommand \bibnamefont  [1]{#1}%
\providecommand \bibfnamefont [1]{#1}%
\providecommand \citenamefont [1]{#1}%
\providecommand \href@noop [0]{\@secondoftwo}%
\providecommand \href [0]{\begingroup \@sanitize@url \@href}%
\providecommand \@href[1]{\@@startlink{#1}\@@href}%
\providecommand \@@href[1]{\endgroup#1\@@endlink}%
\providecommand \@sanitize@url [0]{\catcode `\\12\catcode `\$12\catcode
  `\&12\catcode `\#12\catcode `\^12\catcode `\_12\catcode `\%12\relax}%
\providecommand \@@startlink[1]{}%
\providecommand \@@endlink[0]{}%
\providecommand \url  [0]{\begingroup\@sanitize@url \@url }%
\providecommand \@url [1]{\endgroup\@href {#1}{\urlprefix }}%
\providecommand \urlprefix  [0]{URL }%
\providecommand \Eprint [0]{\href }%
\providecommand \doibase [0]{http://dx.doi.org/}%
\providecommand \selectlanguage [0]{\@gobble}%
\providecommand \bibinfo  [0]{\@secondoftwo}%
\providecommand \bibfield  [0]{\@secondoftwo}%
\providecommand \translation [1]{[#1]}%
\providecommand \BibitemOpen [0]{}%
\providecommand \bibitemStop [0]{}%
\providecommand \bibitemNoStop [0]{.\EOS\space}%
\providecommand \EOS [0]{\spacefactor3000\relax}%
\providecommand \BibitemShut  [1]{\csname bibitem#1\endcsname}%
\let\auto@bib@innerbib\@empty
\bibitem [{\citenamefont {Helstrom}(1976)}]{Helstrom:1976ij}%
  \BibitemOpen
  \bibfield  {author} {\bibinfo {author} {\bibfnamefont {C.~W.}\ \bibnamefont
  {Helstrom}},\ }\href@noop {} {\emph {\bibinfo {title} {Quantum {D}etection
  and {E}stimation {T}heory}}}\ (\bibinfo  {publisher} {Academic},\ \bibinfo
  {address} {New York},\ \bibinfo {year} {1976})\BibitemShut {NoStop}%
\bibitem [{\citenamefont {Holevo}(2003)}]{Holevo:2003fv}%
  \BibitemOpen
  \bibfield  {author} {\bibinfo {author} {\bibfnamefont {A.~S.}\ \bibnamefont
  {Holevo}},\ }\href@noop {} {\emph {\bibinfo {title} {Probabilistic and
  {S}tatistical {A}spects of {Q}uantum {T}heory}}},\ \bibinfo {edition} {2nd}\
  ed.\ (\bibinfo  {publisher} {North Holland},\ \bibinfo {address}
  {Amsterdam},\ \bibinfo {year} {2003})\BibitemShut {NoStop}%
\bibitem [{\citenamefont {Giovannetti}\ \emph {et~al.}(2011)\citenamefont
  {Giovannetti}, \citenamefont {Lloyd},\ and\ \citenamefont
  {Maccone}}]{Giovannetti:2011aa}%
  \BibitemOpen
  \bibfield  {author} {\bibinfo {author} {\bibfnamefont {V.}~\bibnamefont
  {Giovannetti}}, \bibinfo {author} {\bibfnamefont {S.}~\bibnamefont {Lloyd}},
  \ and\ \bibinfo {author} {\bibfnamefont {L.}~\bibnamefont {Maccone}},\
  }\bibfield  {title} {\enquote {\bibinfo {title} {Advances in quantum
  metrology},}\ }\href {http://dx.doi.org/10.1038/nphoton.2011.35} {\bibfield
  {journal} {\bibinfo  {journal} {Nat. Photon.}\ }\textbf {\bibinfo {volume}
  {5}},\ \bibinfo {pages} {222--229} (\bibinfo {year} {2011})}\BibitemShut
  {NoStop}%
\bibitem [{\citenamefont {Demkowicz-Dobrzanski}\ \emph
  {et~al.}(2015)\citenamefont {Demkowicz-Dobrzanski}, \citenamefont {Jarzyna},\
  and\ \citenamefont {Kolodynski}}]{Demkowicz:2015aa}%
  \BibitemOpen
  \bibfield  {author} {\bibinfo {author} {\bibfnamefont {R.}~\bibnamefont
  {Demkowicz-Dobrzanski}}, \bibinfo {author} {\bibfnamefont {M.}~\bibnamefont
  {Jarzyna}}, \ and\ \bibinfo {author} {\bibfnamefont {J.}~\bibnamefont
  {Kolodynski}},\ }\bibfield  {title} {\enquote {\bibinfo {title} {Quantum
  limits in optical interferometry},}\ }\href {\doibase
  https://doi.org/10.1016/bs.po.2015.02.003} {\bibfield  {journal} {\bibinfo
  {journal} {Prog. Opt.}\ }\textbf {\bibinfo {volume} {60}},\ \bibinfo {pages}
  {345--435} (\bibinfo {year} {2015})}\BibitemShut {NoStop}%
\bibitem [{\citenamefont {Braunstein}\ and\ \citenamefont
  {Caves}(1994)}]{Braunstein:1994aa}%
  \BibitemOpen
  \bibfield  {author} {\bibinfo {author} {\bibfnamefont {S.~L.}\ \bibnamefont
  {Braunstein}}\ and\ \bibinfo {author} {\bibfnamefont {C.~M.}\ \bibnamefont
  {Caves}},\ }\bibfield  {title} {\enquote {\bibinfo {title} {Statistical
  distance and the geometry of quantum states},}\ }\href
  {http://link.aps.org/doi/10.1103/PhysRevLett.72.3439} {\bibfield  {journal}
  {\bibinfo  {journal} {Phys. Rev. Lett.}\ }\textbf {\bibinfo {volume} {72}},\
  \bibinfo {pages} {3439--3443} (\bibinfo {year} {1994})}\BibitemShut {NoStop}%
\bibitem [{\citenamefont {Yuen}\ and\ \citenamefont {Lax}(1973)}]{Yuen:1973aa}%
  \BibitemOpen
  \bibfield  {author} {\bibinfo {author} {\bibfnamefont {H.}~\bibnamefont
  {Yuen}}\ and\ \bibinfo {author} {\bibfnamefont {M.}~\bibnamefont {Lax}},\
  }\bibfield  {title} {\enquote {\bibinfo {title} {Multiple parameter quantum
  estimation and measurement of nonselfadjoint observables},}\ }\href@noop {}
  {\bibfield  {journal} {\bibinfo  {journal} {IEEE Trans. Inf. Theory}\
  }\textbf {\bibinfo {volume} {19}},\ \bibinfo {pages} {740--750} (\bibinfo
  {year} {1973})}\BibitemShut {NoStop}%
\bibitem [{\citenamefont {Belavkin}(1976)}]{Belavkin:1976aa}%
  \BibitemOpen
  \bibfield  {author} {\bibinfo {author} {\bibfnamefont {V.~P.}\ \bibnamefont
  {Belavkin}},\ }\bibfield  {title} {\enquote {\bibinfo {title} {Generalized
  uncertainty relations and efficient measurements in quantum systems},}\
  }\href {\doibase 10.1007/BF01032091} {\bibfield  {journal} {\bibinfo
  {journal} {Theor. Math. Phys.}\ }\textbf {\bibinfo {volume} {26}},\ \bibinfo
  {pages} {213--222} (\bibinfo {year} {1976})}\BibitemShut {NoStop}%
\bibitem [{\citenamefont {Szczykulska}\ \emph {et~al.}(2016)\citenamefont
  {Szczykulska}, \citenamefont {Baumgratz},\ and\ \citenamefont
  {Datta}}]{Szczykulska:2016aa}%
  \BibitemOpen
  \bibfield  {author} {\bibinfo {author} {\bibfnamefont {M.}~\bibnamefont
  {Szczykulska}}, \bibinfo {author} {\bibfnamefont {T.}~\bibnamefont
  {Baumgratz}}, \ and\ \bibinfo {author} {\bibfnamefont {A.}~\bibnamefont
  {Datta}},\ }\bibfield  {title} {\enquote {\bibinfo {title} {Multi-parameter
  quantum metrology},}\ }\href {\doibase 10.1080/23746149.2016.1230476}
  {\bibfield  {journal} {\bibinfo  {journal} {Adv. Phys. X}\ }\textbf {\bibinfo
  {volume} {1}},\ \bibinfo {pages} {621--639} (\bibinfo {year}
  {2016})}\BibitemShut {NoStop}%
\bibitem [{\citenamefont {Helstrom}(1967)}]{Helstrom:1967aa}%
  \BibitemOpen
  \bibfield  {author} {\bibinfo {author} {\bibfnamefont {C.~W.}\ \bibnamefont
  {Helstrom}},\ }\bibfield  {title} {\enquote {\bibinfo {title} {Minimum
  mean-squared error of estimates in quantum statistics},}\ }\href {\doibase
  https://doi.org/10.1016/0375-9601(67)90366-0} {\bibfield  {journal} {\bibinfo
   {journal} {Phys. Lett. A}\ }\textbf {\bibinfo {volume} {25}},\ \bibinfo
  {pages} {101--102} (\bibinfo {year} {1967})}\BibitemShut {NoStop}%
\bibitem [{\citenamefont {Matsumoto}(2002)}]{Matsumoto:2002aa}%
  \BibitemOpen
  \bibfield  {author} {\bibinfo {author} {\bibfnamefont {K.}~\bibnamefont
  {Matsumoto}},\ }\bibfield  {title} {\enquote {\bibinfo {title} {A new
  approach to the {C}ram{\'e}r-{R}ao-type bound of the pure-state model},}\
  }\href {http://stacks.iop.org/0305-4470/35/i=13/a=307} {\bibfield  {journal}
  {\bibinfo  {journal} {J. Phys. A: Math. Gen.}\ }\textbf {\bibinfo {volume}
  {35}},\ \bibinfo {pages} {3111} (\bibinfo {year} {2002})}\BibitemShut
  {NoStop}%
\bibitem [{\citenamefont {Gill}\ and\ \citenamefont
  {Guta}(2013)}]{Gill:2013aa}%
  \BibitemOpen
  \bibfield  {author} {\bibinfo {author} {\bibfnamefont {R.~D.}\ \bibnamefont
  {Gill}}\ and\ \bibinfo {author} {\bibfnamefont {M.~I.}\ \bibnamefont
  {Guta}},\ }\enquote {\bibinfo {title} {On asymptotic quantum statistical
  inference},}\ \ (\bibinfo  {publisher} {Institute of Mathematical
  Statistics},\ \bibinfo {address} {Beachwood, Ohio, USA},\ \bibinfo {year}
  {2013})\ pp.\ \bibinfo {pages} {105--127}\BibitemShut {NoStop}%
\bibitem [{\citenamefont {Pezz{\`e}}\ \emph {et~al.}(2017)\citenamefont
  {Pezz{\`e}}, \citenamefont {Ciampini}, \citenamefont {Spagnolo},
  \citenamefont {Humphreys}, \citenamefont {Datta}, \citenamefont {Walmsley},
  \citenamefont {Barbieri}, \citenamefont {Sciarrino},\ and\ \citenamefont
  {Smerzi}}]{Pezze:2017aa}%
  \BibitemOpen
  \bibfield  {author} {\bibinfo {author} {\bibfnamefont {L.}~\bibnamefont
  {Pezz{\`e}}}, \bibinfo {author} {\bibfnamefont {M.~A.}\ \bibnamefont
  {Ciampini}}, \bibinfo {author} {\bibfnamefont {N.}~\bibnamefont {Spagnolo}},
  \bibinfo {author} {\bibfnamefont {P.~C.}\ \bibnamefont {Humphreys}}, \bibinfo
  {author} {\bibfnamefont {A.}~\bibnamefont {Datta}}, \bibinfo {author}
  {\bibfnamefont {I.~A.}\ \bibnamefont {Walmsley}}, \bibinfo {author}
  {\bibfnamefont {M.}~\bibnamefont {Barbieri}}, \bibinfo {author}
  {\bibfnamefont {F}~\bibnamefont {Sciarrino}}, \ and\ \bibinfo {author}
  {\bibfnamefont {A.}~\bibnamefont {Smerzi}},\ }\bibfield  {title} {\enquote
  {\bibinfo {title} {Optimal measurements for simultaneous quantum estimation
  of multiple phases},}\ }\href
  {https://link.aps.org/doi/10.1103/PhysRevLett.119.130504} {\bibfield
  {journal} {\bibinfo  {journal} {Phys. Rev. Lett.}\ }\textbf {\bibinfo
  {volume} {119}},\ \bibinfo {pages} {130504} (\bibinfo {year}
  {2017})}\BibitemShut {NoStop}%
\bibitem [{\citenamefont {Yue}\ \emph {et~al.}(2014)\citenamefont {Yue},
  \citenamefont {Zhang},\ and\ \citenamefont {Fan}}]{Yue:2014aa}%
  \BibitemOpen
  \bibfield  {author} {\bibinfo {author} {\bibfnamefont {J.-D.}\ \bibnamefont
  {Yue}}, \bibinfo {author} {\bibfnamefont {Y.-R.}\ \bibnamefont {Zhang}}, \
  and\ \bibinfo {author} {\bibfnamefont {H.}~\bibnamefont {Fan}},\ }\bibfield
  {title} {\enquote {\bibinfo {title} {Quantum-enhanced metrology for multiple
  phase estimation with noise},}\ }\href {http://dx.doi.org/10.1038/srep05933}
  {\bibfield  {journal} {\bibinfo  {journal} {Sci. Rep.}\ }\textbf {\bibinfo
  {volume} {4}},\ \bibinfo {pages} {5933 EP} (\bibinfo {year}
  {2014})}\BibitemShut {NoStop}%
\bibitem [{\citenamefont {Zuppardo}\ \emph {et~al.}(2015)\citenamefont
  {Zuppardo}, \citenamefont {Santos}, \citenamefont {De~Chiara}, \citenamefont
  {Paternostro}, \citenamefont {Semi{\~a}o},\ and\ \citenamefont
  {Palma}}]{Zuppardo:2015aa}%
  \BibitemOpen
  \bibfield  {author} {\bibinfo {author} {\bibfnamefont {M.}~\bibnamefont
  {Zuppardo}}, \bibinfo {author} {\bibfnamefont {J.~P.}\ \bibnamefont
  {Santos}}, \bibinfo {author} {\bibfnamefont {G.}~\bibnamefont {De~Chiara}},
  \bibinfo {author} {\bibfnamefont {M.}~\bibnamefont {Paternostro}}, \bibinfo
  {author} {\bibfnamefont {F.~L.}\ \bibnamefont {Semi{\~a}o}}, \ and\ \bibinfo
  {author} {\bibfnamefont {G.~M.}\ \bibnamefont {Palma}},\ }\bibfield  {title}
  {\enquote {\bibinfo {title} {Cavity-aided quantum parameter estimation in a
  bosonic double-well {J}osephson junction},}\ }\href
  {https://link.aps.org/doi/10.1103/PhysRevA.91.033631} {\bibfield  {journal}
  {\bibinfo  {journal} {Phys. Rev. A}\ }\textbf {\bibinfo {volume} {91}},\
  \bibinfo {pages} {033631} (\bibinfo {year} {2015})}\BibitemShut {NoStop}%
\bibitem [{\citenamefont {Proctor}\ \emph {et~al.}(2018)\citenamefont
  {Proctor}, \citenamefont {Knott},\ and\ \citenamefont
  {Dunningham}}]{Proctor:2018aa}%
  \BibitemOpen
  \bibfield  {author} {\bibinfo {author} {\bibfnamefont {T.~J.}\ \bibnamefont
  {Proctor}}, \bibinfo {author} {\bibfnamefont {P.~A.}\ \bibnamefont {Knott}},
  \ and\ \bibinfo {author} {\bibfnamefont {J.~A.}\ \bibnamefont {Dunningham}},\
  }\bibfield  {title} {\enquote {\bibinfo {title} {Multiparameter estimation in
  networked quantum sensors},}\ }\href
  {https://link.aps.org/doi/10.1103/PhysRevLett.120.080501} {\bibfield
  {journal} {\bibinfo  {journal} {Phys. Rev. Lett.}\ }\textbf {\bibinfo
  {volume} {120}},\ \bibinfo {pages} {080501} (\bibinfo {year}
  {2018})}\BibitemShut {NoStop}%
\bibitem [{\citenamefont {den Dekker}\ and\ \citenamefont {van~den
  Bos}(1997)}]{Dekker:1997aa}%
  \BibitemOpen
  \bibfield  {author} {\bibinfo {author} {\bibfnamefont {A.~J.}\ \bibnamefont
  {den Dekker}}\ and\ \bibinfo {author} {\bibfnamefont {A.}~\bibnamefont
  {van~den Bos}},\ }\bibfield  {title} {\enquote {\bibinfo {title} {Resolution:
  a survey},}\ }\href {\doibase 10.1364/JOSAA.14.000547} {\bibfield  {journal}
  {\bibinfo  {journal} {J. Opt. Soc. Am. A}\ }\textbf {\bibinfo {volume}
  {14}},\ \bibinfo {pages} {547--557} (\bibinfo {year} {1997})}\BibitemShut
  {NoStop}%
\bibitem [{\citenamefont {Hemmer}\ and\ \citenamefont
  {Zapata}(2012)}]{Hemmer:2012aa}%
  \BibitemOpen
  \bibfield  {author} {\bibinfo {author} {\bibfnamefont {P.~R.}\ \bibnamefont
  {Hemmer}}\ and\ \bibinfo {author} {\bibfnamefont {T.}~\bibnamefont
  {Zapata}},\ }\bibfield  {title} {\enquote {\bibinfo {title} {The universal
  scaling laws that determine the achievable resolution in different schemes
  for super-resolution imaging},}\ }\href
  {http://stacks.iop.org/2040-8986/14/i=8/a=083002} {\bibfield  {journal}
  {\bibinfo  {journal} {J. Opt.}\ }\textbf {\bibinfo {volume} {14}},\ \bibinfo
  {pages} {083002} (\bibinfo {year} {2012})}\BibitemShut {NoStop}%
\bibitem [{\citenamefont {Cremer}\ and\ \citenamefont
  {Masters}(2013)}]{Cremer:2013aa}%
  \BibitemOpen
  \bibfield  {author} {\bibinfo {author} {\bibfnamefont {C.}~\bibnamefont
  {Cremer}}\ and\ \bibinfo {author} {\bibfnamefont {R.~B.}\ \bibnamefont
  {Masters}},\ }\bibfield  {title} {\enquote {\bibinfo {title} {Resolution
  enhancement techniques in microscopy},}\ }\href {\doibase
  10.1140/epjh/e2012-20060-1} {\bibfield  {journal} {\bibinfo  {journal} {Eur.
  Phys. J. H}\ }\textbf {\bibinfo {volume} {38}},\ \bibinfo {pages} {281--344}
  (\bibinfo {year} {2013})}\BibitemShut {NoStop}%
\bibitem [{\citenamefont {Rayleigh}(1879)}]{Rayleigh:1879aa}%
  \BibitemOpen
  \bibfield  {author} {\bibinfo {author} {\bibfnamefont {Lord}\ \bibnamefont
  {Rayleigh}},\ }\bibfield  {title} {\enquote {\bibinfo {title} {Investigations
  in {O}ptics, with special reference to the spectroscope},}\ }\href@noop {}
  {\bibfield  {journal} {\bibinfo  {journal} {Phil. Mag.}\ }\textbf {\bibinfo
  {volume} {8}},\ \bibinfo {pages} {261--274, 403--411, 477--486} (\bibinfo
  {year} {1879})}\BibitemShut {NoStop}%
\bibitem [{\citenamefont {Schermelleh}\ \emph {et~al.}(2010)\citenamefont
  {Schermelleh}, \citenamefont {Heintzmann},\ and\ \citenamefont
  {Leonhardt}}]{Schermelleh:2010aa}%
  \BibitemOpen
  \bibfield  {author} {\bibinfo {author} {\bibfnamefont {L.}~\bibnamefont
  {Schermelleh}}, \bibinfo {author} {\bibfnamefont {R.}~\bibnamefont
  {Heintzmann}}, \ and\ \bibinfo {author} {\bibfnamefont {H.}~\bibnamefont
  {Leonhardt}},\ }\bibfield  {title} {\enquote {\bibinfo {title} {A guide to
  super-resolution fluorescence microscopy},}\ }\href
  {http://jcb.rupress.org/content/190/2/165.abstract} {\bibfield  {journal}
  {\bibinfo  {journal} {J. Cell Biol.}\ }\textbf {\bibinfo {volume} {190}},\
  \bibinfo {pages} {165} (\bibinfo {year} {2010})}\BibitemShut {NoStop}%
\bibitem [{\citenamefont {Tsang}\ \emph {et~al.}(2016)\citenamefont {Tsang},
  \citenamefont {Nair},\ and\ \citenamefont {Lu}}]{Tsang:2016aa}%
  \BibitemOpen
  \bibfield  {author} {\bibinfo {author} {\bibfnamefont {M.}~\bibnamefont
  {Tsang}}, \bibinfo {author} {\bibfnamefont {R.}~\bibnamefont {Nair}}, \ and\
  \bibinfo {author} {\bibfnamefont {X.-M.}\ \bibnamefont {Lu}},\ }\bibfield
  {title} {\enquote {\bibinfo {title} {Quantum theory of superresolution for
  two incoherent optical point sources},}\ }\href
  {http://link.aps.org/doi/10.1103/PhysRevX.6.031033} {\bibfield  {journal}
  {\bibinfo  {journal} {Phys. Rev.~X}\ }\textbf {\bibinfo {volume} {6}},\
  \bibinfo {pages} {031033} (\bibinfo {year} {2016})}\BibitemShut {NoStop}%
\bibitem [{\citenamefont {Nair}\ and\ \citenamefont
  {Tsang}(2016)}]{Nair:2016aa}%
  \BibitemOpen
  \bibfield  {author} {\bibinfo {author} {\bibfnamefont {R.}~\bibnamefont
  {Nair}}\ and\ \bibinfo {author} {\bibfnamefont {M.}~\bibnamefont {Tsang}},\
  }\bibfield  {title} {\enquote {\bibinfo {title} {Far-field superresolution of
  thermal electromagnetic sources at the quantum limit},}\ }\href
  {http://arxiv.org/abs/1604.00937} {\bibfield  {journal} {\bibinfo  {journal}
  {Phys. Rev. Lett.}\ }\textbf {\bibinfo {volume} {117}},\ \bibinfo {pages}
  {190801} (\bibinfo {year} {2016})}\BibitemShut {NoStop}%
\bibitem [{\citenamefont {Ang}\ \emph {et~al.}(2016)\citenamefont {Ang},
  \citenamefont {Nair},\ and\ \citenamefont {Tsang}}]{Ang:2016aa}%
  \BibitemOpen
  \bibfield  {author} {\bibinfo {author} {\bibfnamefont {S.~Z.}\ \bibnamefont
  {Ang}}, \bibinfo {author} {\bibfnamefont {R.}~\bibnamefont {Nair}}, \ and\
  \bibinfo {author} {\bibfnamefont {M.}~\bibnamefont {Tsang}},\ }\bibfield
  {title} {\enquote {\bibinfo {title} {Quantum limit for two-dimensional
  resolution of two incoherent optical point sources},}\ }\href
  {http://arxiv.org/pdf/1606.00603.pdf} {\bibfield  {journal} {\bibinfo
  {journal} {Phys. Rev. A}\ }\textbf {\bibinfo {volume} {95}},\ \bibinfo
  {pages} {063847} (\bibinfo {year} {2016})}\BibitemShut {NoStop}%
\bibitem [{\citenamefont {Lupo}\ and\ \citenamefont
  {Pirandola}(2016)}]{Lupo:2016aa}%
  \BibitemOpen
  \bibfield  {author} {\bibinfo {author} {\bibfnamefont {C.}~\bibnamefont
  {Lupo}}\ and\ \bibinfo {author} {\bibfnamefont {S.}~\bibnamefont
  {Pirandola}},\ }\bibfield  {title} {\enquote {\bibinfo {title} {Ultimate
  precision bound of quantum and subwavelength imaging},}\ }\href
  {http://arxiv.org/abs/1604.07367} {\bibfield  {journal} {\bibinfo  {journal}
  {Phys. Rev. Lett.}\ }\textbf {\bibinfo {volume} {117}},\ \bibinfo {pages}
  {190802} (\bibinfo {year} {2016})}\BibitemShut {NoStop}%
\bibitem [{\citenamefont {Rehacek}\ \emph
  {et~al.}(2017{\natexlab{a}})\citenamefont {Rehacek}, \citenamefont
  {Pa{\'u}r}, \citenamefont {Stoklasa}, \citenamefont {Hradil},\ and\
  \citenamefont {S{\'a}nchez-Soto}}]{Rehacek:2017aa}%
  \BibitemOpen
  \bibfield  {author} {\bibinfo {author} {\bibfnamefont {J.}~\bibnamefont
  {Rehacek}}, \bibinfo {author} {\bibfnamefont {M.}~\bibnamefont {Pa{\'u}r}},
  \bibinfo {author} {\bibfnamefont {B.}~\bibnamefont {Stoklasa}}, \bibinfo
  {author} {\bibfnamefont {Z.}~\bibnamefont {Hradil}}, \ and\ \bibinfo {author}
  {\bibfnamefont {L.~L.}\ \bibnamefont {S{\'a}nchez-Soto}},\ }\bibfield
  {title} {\enquote {\bibinfo {title} {Optimal measurements for resolution
  beyond the {R}ayleigh limit},}\ }\href {\doibase 10.1364/OL.42.000231}
  {\bibfield  {journal} {\bibinfo  {journal} {Opt. Lett.}\ }\textbf {\bibinfo
  {volume} {42}},\ \bibinfo {pages} {231--234} (\bibinfo {year}
  {2017}{\natexlab{a}})}\BibitemShut {NoStop}%
\bibitem [{\citenamefont {Kerviche}\ \emph {et~al.}(2017)\citenamefont
  {Kerviche}, \citenamefont {Guha},\ and\ \citenamefont
  {Ashok}}]{Kerviche:2017aa}%
  \BibitemOpen
  \bibfield  {author} {\bibinfo {author} {\bibfnamefont {R.}~\bibnamefont
  {Kerviche}}, \bibinfo {author} {\bibfnamefont {S.}~\bibnamefont {Guha}}, \
  and\ \bibinfo {author} {\bibfnamefont {A.}~\bibnamefont {Ashok}},\ }\bibfield
   {title} {\enquote {\bibinfo {title} {Fundamental limit of resolving two
  point sources limited by an arbitrary point spread function},}\ }in\ \href
  {\doibase 10.1109/ISIT.2017.8006566} {\emph {\bibinfo {booktitle} {2017 IEEE
  International Symposium on Information Theory (ISIT)}}}\ (\bibinfo {year}
  {2017})\ pp.\ \bibinfo {pages} {441--445}\BibitemShut {NoStop}%
\bibitem [{\citenamefont {Tsang}(2018)}]{Tsang:2018aa}%
  \BibitemOpen
  \bibfield  {author} {\bibinfo {author} {\bibfnamefont {M.}~\bibnamefont
  {Tsang}},\ }\bibfield  {title} {\enquote {\bibinfo {title} {Conservative
  classical and quantum resolution limits for incoherent imaging},}\ }\href
  {\doibase 10.1080/09500340.2017.1377306} {\bibfield  {journal} {\bibinfo
  {journal} {J. Mod. Opt.}\ }\textbf {\bibinfo {volume} {65}},\ \bibinfo
  {pages} {104--110} (\bibinfo {year} {2018})}\BibitemShut {NoStop}%
\bibitem [{\citenamefont {Zhou}\ and\ \citenamefont
  {Jiang}(2018)}]{Zhou:2018aa}%
  \BibitemOpen
  \bibfield  {author} {\bibinfo {author} {\bibfnamefont {S.}~\bibnamefont
  {Zhou}}\ and\ \bibinfo {author} {\bibfnamefont {L.}~\bibnamefont {Jiang}},\
  }\bibfield  {title} {\enquote {\bibinfo {title} {A modern description of
  {R}ayleigh's criterion},}\ }\href@noop {} {\bibfield  {journal} {\bibinfo
  {journal} {preprint arXiv:1801.02917}\ } (\bibinfo {year}
  {2018})}\BibitemShut {NoStop}%
\bibitem [{\citenamefont {Paur}\ \emph {et~al.}(2016)\citenamefont {Paur},
  \citenamefont {Stoklasa}, \citenamefont {Hradil}, \citenamefont
  {Sanchez-Soto},\ and\ \citenamefont {Rehacek}}]{Paur:2016aa}%
  \BibitemOpen
  \bibfield  {author} {\bibinfo {author} {\bibfnamefont {M.}~\bibnamefont
  {Paur}}, \bibinfo {author} {\bibfnamefont {B.}~\bibnamefont {Stoklasa}},
  \bibinfo {author} {\bibfnamefont {Z.}~\bibnamefont {Hradil}}, \bibinfo
  {author} {\bibfnamefont {L.~L.}\ \bibnamefont {Sanchez-Soto}}, \ and\
  \bibinfo {author} {\bibfnamefont {J.}~\bibnamefont {Rehacek}},\ }\bibfield
  {title} {\enquote {\bibinfo {title} {Achieving the ultimate optical
  resolution},}\ }\href {\doibase 10.1364/OPTICA.3.001144} {\bibfield
  {journal} {\bibinfo  {journal} {Optica}\ }\textbf {\bibinfo {volume} {3}},\
  \bibinfo {pages} {1144--1147} (\bibinfo {year} {2016})}\BibitemShut {NoStop}%
\bibitem [{\citenamefont {Yang}\ \emph {et~al.}(2016)\citenamefont {Yang},
  \citenamefont {Taschilina}, \citenamefont {Moiseev}, \citenamefont {Simon},\
  and\ \citenamefont {Lvovsky}}]{Yang:2016aa}%
  \BibitemOpen
  \bibfield  {author} {\bibinfo {author} {\bibfnamefont {F.}~\bibnamefont
  {Yang}}, \bibinfo {author} {\bibfnamefont {A.}~\bibnamefont {Taschilina}},
  \bibinfo {author} {\bibfnamefont {E.~S.}\ \bibnamefont {Moiseev}}, \bibinfo
  {author} {\bibfnamefont {C.}~\bibnamefont {Simon}}, \ and\ \bibinfo {author}
  {\bibfnamefont {A.~I.}\ \bibnamefont {Lvovsky}},\ }\bibfield  {title}
  {\enquote {\bibinfo {title} {Far-field linear optical superresolution via
  heterodyne detection in a higher-order local oscillator mode},}\ }\href
  {\doibase 10.1364/OPTICA.3.001148} {\bibfield  {journal} {\bibinfo  {journal}
  {Optica}\ }\textbf {\bibinfo {volume} {3}},\ \bibinfo {pages} {1148--1152}
  (\bibinfo {year} {2016})}\BibitemShut {NoStop}%
\bibitem [{\citenamefont {Tham}\ \emph {et~al.}(2016)\citenamefont {Tham},
  \citenamefont {Ferretti},\ and\ \citenamefont {Steinberg}}]{Tham:2016aa}%
  \BibitemOpen
  \bibfield  {author} {\bibinfo {author} {\bibfnamefont {W.~K.}\ \bibnamefont
  {Tham}}, \bibinfo {author} {\bibfnamefont {H.}~\bibnamefont {Ferretti}}, \
  and\ \bibinfo {author} {\bibfnamefont {A.~M.}\ \bibnamefont {Steinberg}},\
  }\bibfield  {title} {\enquote {\bibinfo {title} {Beating {R}ayleigh's curse
  by imaging using phase information},}\ }\href
  {http://arxiv.org/pdf/1606.02666.pdf} {\bibfield  {journal} {\bibinfo
  {journal} {Phys. Rev. Lett.}\ }\textbf {\bibinfo {volume} {118}},\ \bibinfo
  {pages} {070801} (\bibinfo {year} {2016})}\BibitemShut {NoStop}%
\bibitem [{\citenamefont {Yang}\ \emph {et~al.}(2017)\citenamefont {Yang},
  \citenamefont {Nair}, \citenamefont {Tsang}, \citenamefont {Simon},\ and\
  \citenamefont {Lvovsky}}]{Yang:2017aa}%
  \BibitemOpen
  \bibfield  {author} {\bibinfo {author} {\bibfnamefont {F.}~\bibnamefont
  {Yang}}, \bibinfo {author} {\bibfnamefont {R.}~\bibnamefont {Nair}}, \bibinfo
  {author} {\bibfnamefont {M.}~\bibnamefont {Tsang}}, \bibinfo {author}
  {\bibfnamefont {C.}~\bibnamefont {Simon}}, \ and\ \bibinfo {author}
  {\bibfnamefont {A.~I.}\ \bibnamefont {Lvovsky}},\ }\bibfield  {title}
  {\enquote {\bibinfo {title} {Fisher information for far-field linear optical
  superresolution via homodyne or heterodyne detection in a higher-order local
  oscillator mode},}\ }\href {\doibase https://arxiv.org/pdf/1706.08633.pdf}
  {\bibfield  {journal} {\bibinfo  {journal} {Phys. Rev. A}\ }\textbf {\bibinfo
  {volume} {96}},\ \bibinfo {pages} {063829} (\bibinfo {year}
  {2017})}\BibitemShut {NoStop}%
\bibitem [{\citenamefont {Chrostowski}\ \emph {et~al.}(2017)\citenamefont
  {Chrostowski}, \citenamefont {Demkowicz-Dobrza\'{n}ski}, \citenamefont
  {Jarzyna},\ and\ \citenamefont {Banaszek}}]{Chrostowski:2017aa}%
  \BibitemOpen
  \bibfield  {author} {\bibinfo {author} {\bibfnamefont {A.}~\bibnamefont
  {Chrostowski}}, \bibinfo {author} {\bibfnamefont {R.}~\bibnamefont
  {Demkowicz-Dobrza\'{n}ski}}, \bibinfo {author} {\bibfnamefont
  {M.}~\bibnamefont {Jarzyna}}, \ and\ \bibinfo {author} {\bibfnamefont
  {K.}~\bibnamefont {Banaszek}},\ }\bibfield  {title} {\enquote {\bibinfo
  {title} {On superresolution imaging as a multiparameter estimation
  problem},}\ }\href {\doibase 10.1142/S0219749917400056} {\bibfield  {journal}
  {\bibinfo  {journal} {Int. J. Quantum Inf.}\ }\textbf {\bibinfo {volume}
  {17}},\ \bibinfo {pages} {1740005} (\bibinfo {year} {2017})}\BibitemShut
  {NoStop}%
\bibitem [{\citenamefont {Rehacek}\ \emph
  {et~al.}(2017{\natexlab{b}})\citenamefont {Rehacek}, \citenamefont {Hradil},
  \citenamefont {Stoklasa}, \citenamefont {Pa{\'u}r}, \citenamefont {Grover},
  \citenamefont {Krzic},\ and\ \citenamefont
  {S{\'a}nchez-Soto}}]{Rehacek:2017ab}%
  \BibitemOpen
  \bibfield  {author} {\bibinfo {author} {\bibfnamefont {J.}~\bibnamefont
  {Rehacek}}, \bibinfo {author} {\bibfnamefont {Z.}~\bibnamefont {Hradil}},
  \bibinfo {author} {\bibfnamefont {B.}~\bibnamefont {Stoklasa}}, \bibinfo
  {author} {\bibfnamefont {M.}~\bibnamefont {Pa{\'u}r}}, \bibinfo {author}
  {\bibfnamefont {J.}~\bibnamefont {Grover}}, \bibinfo {author} {\bibfnamefont
  {A.}~\bibnamefont {Krzic}}, \ and\ \bibinfo {author} {\bibfnamefont {L.~L.}\
  \bibnamefont {S{\'a}nchez-Soto}},\ }\bibfield  {title} {\enquote {\bibinfo
  {title} {Multiparameter quantum metrology of incoherent point sources:
  towards realistic superresolution},}\ }\href {\doibase
  10.1103/PhysRevA.96.062107} {\bibfield  {journal} {\bibinfo  {journal} {Phys.
  Rev. A}\ }\textbf {\bibinfo {volume} {96}},\ \bibinfo {pages} {062107}
  (\bibinfo {year} {2017}{\natexlab{b}})}\BibitemShut {NoStop}%
\bibitem [{\citenamefont {Petz}\ and\ \citenamefont
  {Ghinea}(2011)}]{Petz:2011aa}%
  \BibitemOpen
  \bibfield  {author} {\bibinfo {author} {\bibfnamefont {D.}~\bibnamefont
  {Petz}}\ and\ \bibinfo {author} {\bibfnamefont {C.}~\bibnamefont {Ghinea}},\
  }\enquote {\bibinfo {title} {Introduction to {Q}uantum {F}isher
  {I}nformation},}\ in\ \href {\doibase doi:10.1142/9789814338745_0015} {\emph
  {\bibinfo {booktitle} {Quantum Probability and Related Topics}}},\ Vol.\
  \bibinfo {volume} {Volume 27}\ (\bibinfo  {publisher} {World Scientific},\
  \bibinfo {year} {2011})\ pp.\ \bibinfo {pages} {261--281}\BibitemShut
  {NoStop}%
\bibitem [{\citenamefont {Suzuki}(2016)}]{Suzuki:2016aa}%
  \BibitemOpen
  \bibfield  {author} {\bibinfo {author} {\bibfnamefont {J.}~\bibnamefont
  {Suzuki}},\ }\bibfield  {title} {\enquote {\bibinfo {title} {Explicit formula
  for the {H}olevo bound for two-parameter qubit-state estimation problem},}\
  }\href {\doibase 10.1063/1.4945086} {\bibfield  {journal} {\bibinfo
  {journal} {J. Math. Phys.}\ }\textbf {\bibinfo {volume} {57}},\ \bibinfo
  {pages} {042201} (\bibinfo {year} {2016})}\BibitemShut {NoStop}%
\bibitem [{\citenamefont {Vaneph}\ \emph {et~al.}(2013)\citenamefont {Vaneph},
  \citenamefont {Tufarelli},\ and\ \citenamefont {Genoni.}}]{Vaneph:2013aa}%
  \BibitemOpen
  \bibfield  {author} {\bibinfo {author} {\bibfnamefont {C.}~\bibnamefont
  {Vaneph}}, \bibinfo {author} {\bibfnamefont {T.}~\bibnamefont {Tufarelli}}, \
  and\ \bibinfo {author} {\bibfnamefont {M.~G.}\ \bibnamefont {Genoni.}},\
  }\bibfield  {title} {\enquote {\bibinfo {title} {Quantum estimation of a
  two-phase spin rotation},}\ }\href {\doibase
  https://doi.org/10.2478/qmetro-2013-0003} {\bibfield  {journal} {\bibinfo
  {journal} {Quantum Meas. Quantum Metrol.}\ }\textbf {\bibinfo {volume} {1}},\
  \bibinfo {pages} {12--20} (\bibinfo {year} {2013})}\BibitemShut {NoStop}%
\bibitem [{\citenamefont {Crowley}\ \emph {et~al.}(2014)\citenamefont
  {Crowley}, \citenamefont {Datta}, \citenamefont {Barbieri},\ and\
  \citenamefont {Walmsley}}]{Crowley:2014aa}%
  \BibitemOpen
  \bibfield  {author} {\bibinfo {author} {\bibfnamefont {P.~J.~D.}\
  \bibnamefont {Crowley}}, \bibinfo {author} {\bibfnamefont {A.}~\bibnamefont
  {Datta}}, \bibinfo {author} {\bibfnamefont {M.}~\bibnamefont {Barbieri}}, \
  and\ \bibinfo {author} {\bibfnamefont {I.~A.}\ \bibnamefont {Walmsley}},\
  }\bibfield  {title} {\enquote {\bibinfo {title} {Tradeoff in simultaneous
  quantum-limited phase and loss estimation in interferometry},}\ }\href
  {https://link.aps.org/doi/10.1103/PhysRevA.89.023845} {\bibfield  {journal}
  {\bibinfo  {journal} {Phys. Rev. A}\ }\textbf {\bibinfo {volume} {89}},\
  \bibinfo {pages} {023845} (\bibinfo {year} {2014})}\BibitemShut {NoStop}%
\bibitem [{\citenamefont {Ragy}\ \emph {et~al.}(2016)\citenamefont {Ragy},
  \citenamefont {Jarzyna},\ and\ \citenamefont
  {Demkowicz-Dobrza{\'n}ski}}]{Ragy:2016aa}%
  \BibitemOpen
  \bibfield  {author} {\bibinfo {author} {\bibfnamefont {S.}~\bibnamefont
  {Ragy}}, \bibinfo {author} {\bibfnamefont {M.}~\bibnamefont {Jarzyna}}, \
  and\ \bibinfo {author} {\bibfnamefont {R.}~\bibnamefont
  {Demkowicz-Dobrza{\'n}ski}},\ }\bibfield  {title} {\enquote {\bibinfo {title}
  {Compatibility in multiparameter quantum metrology},}\ }\href
  {https://link.aps.org/doi/10.1103/PhysRevA.94.052108} {\bibfield  {journal}
  {\bibinfo  {journal} {Phys. Rev. A}\ }\textbf {\bibinfo {volume} {94}},\
  \bibinfo {pages} {052108} (\bibinfo {year} {2016})}\BibitemShut {NoStop}%
\bibitem [{\citenamefont {Bernardo}\ and\ \citenamefont
  {Smith}(2000)}]{Bernardo:2000aa}%
  \BibitemOpen
  \bibfield  {author} {\bibinfo {author} {\bibfnamefont {J.~M.}\ \bibnamefont
  {Bernardo}}\ and\ \bibinfo {author} {\bibfnamefont {A.~F.~M.}\ \bibnamefont
  {Smith}},\ }\href@noop {} {\emph {\bibinfo {title} {Bayesian {T}heory}}}\
  (\bibinfo  {publisher} {Wiley},\ \bibinfo {address} {Sussex},\ \bibinfo
  {year} {2000})\BibitemShut {NoStop}%
\end{thebibliography}

%

\end{document}